# Spatial Distribution of City Tweets and Their Densities

Bin Jiang[1], Ding Ma[1], Junjun Yin[2], and Mats Sandberg[1]

[1]Faculty of Engineering and Sustainable Development
University of Gävle, SE-801 76 Gävle, Sweden
Email: bin.jiang|ding.ma|mats.sandberg@hig.se

[2]Department of Geography and Geographic Information Science
University of Illinois at Urbana and Champaign, USA
Email: jyn@illinois.edu

**Abstract**
Social media outlets such as Twitter constitute valuable data sources for understanding human activities in the virtual world from a geographic perspective. This paper examines spatial distribution of tweets and densities within cities. The cities refer to natural cities that are automatically aggregated from a country's small street blocks, so called city blocks. We adopted street blocks (rather than census tracts) as the basic geographic units and topological center (rather than geometric center) in order to assess how tweets and densities vary from the center to the peripheral border. We found that, within a city from the center to the periphery, the tweets first increase and then decrease, while the densities decrease in general. These increases and decreases fluctuate dramatically, and differ significantly from those if census tracts are used as the basic geographic units. We also found that the decrease of densities from the center to the periphery is less significant, and even disappears, if an arbitrarily defined city border is adopted. These findings prove that natural cities and their topological centers are better than their counterparts (conventionally defined cities and city centers) for geographic research. Based on this study, we believe that tweet densities can be a good surrogate of population densities. If this belief is proved to be true, social media data could help solve the dispute surrounding exponential or power function of urban population density.

**Keywords:** Big data, natural cities, street blocks, urban density, topological distance

## 1. Introduction

Increasingly popular social media platforms such as Twitter provide valuable data sources for better understanding human activities in the virtual world from a geographic perspetive (e.g., Crandall et al. 2009, Hawelka et al. 2014, Li et al. 2014, Jiang 2015a). The data collected from these social media constitute what we now call big data. Unlike small data, such as those maintained by census and statistical bureaus, which are mainly sampled although population data attempt to cover everyone or an entire population, big data are automatically harvested from a large number of users or locations, through crawling techniques or application programming interfaces (API). Given the large size, big data are often referred to as all (or a large amount of) rather than a small part of the data itself. Unlike small data that are essentially estimated or statistically inferenced, big data such as social media data are accurately measured at very fine spatial and temperal resolutions. Therefore, unlike small data that are aggregated into census tracts or their centroids, big data are defined at an individual level, measured by fine GPS locations and time stamps. These three distinguishing features – all, measured, and individual (Mayer-Schonberger and Cukier 2013) – make social media data a powerful instrument and even a new paradigm for geographic research (Jiang 2015c). For example, the location-based and time-stamped social media data from Brightkite and Twitter can efficiently and effectively illustate structure and dynamics of cities to enhance our understanding of underlying evolution mechanisms (Ferrari et al. 2011, Wakamiya et al. 2011, Cranshaw et al. 2012, Jiang and Miao 2014). The present paper attempts to examine the spatial distribution of tweets and densities within cities. The cities refer to natural cities (Jiang and Liu 2012) that are delineated from a country's all street blocks, which are created by streets, and cycling and pedestrian routes.

The massive number of street blocks of a country, up to two millions, can be partitioned into two parts: those above an average, called field blocks in the countryside, which are a minority, and those below the average, called city blocks in the cities, which are a majority. The city blocks or small street blocks can be merged together to constitute individual patches, which are called natural cities (see Section 2 for the definition and Figure 1 for an illustration). This way of defining cities is essentially bottom up, since all individual blocks collectively determine an average, and then let the average to partition all the street blocks into field and city blocks. In comparison to conventional definitions, often based on some arbitrary threshold, e.g., population > 10,000 as a cutoff, the natural cities are defined objectively or naturally. In other words, conventional definitions are imposed from the top down, while the natural cities emerge from the bottom up. It is important to note that the street blocks, unlike census blocks, are automatically detected from a street network, which can contain cycling and pedestrian paths.

In this study, city blocks are the basic geographic units for examining the spatial distribution of tweets and densities from a city center to its periphery. City blocks are the smallest geographic units identifiable for cities – much smaller than commonly used census tracts or similar. For example, the London natural city studied in this paper contains 45,627 city blocks, but it has only 21,646 output areas, the smallest census unit, and 628 census tracts; in other words, the number of city blocks is twice as many as that of the output areas, and 72 times as many as that of the census tracts. The fine scale of city blocks enables us to study tweets and their densities at a very detailed scale. It should be noted that the city blocks defined from the bottom up have never been used as basic geographic units for geographic research. The fine scale of city blocks also presents the need for us to reassess earlier findings about geographic phenomena such as population density that were studied at some coarse scales.

This study is further motivated by the notion of topological center, which is defined as the city block(s) with the highest border number (see Section 2 for more details or Figure 2 for an illustration). Conventionally, there are multiple ways of defining city center such as historic centers, and central business districts. These definitions are too vague for big data analytics. Instead we must adopt some universal definition like the topological center. The concepts of natural cities, and topological center, and the fine scale of city blocks, as well as related topological and geometric distances, constitute some unique and novel aspects of this study. We attempted to show spatial distribution of tweets and densities at the scale of city blocks, and how the spatial distribution differs if census tracts or an arbitrarily imposed city boundary are adopted. Through the investigation, we demonstrate the effectiveness of natural cities, city blocks, and the topological centers for geographic research in the context of big data.

The remainder of this paper is structured as follows. Section 2 presents the concepts of natural cities, street blocks, and the three different distances: topological, geometric, and radial. Section 3 describes in detail the data and data processing for the tweet location data from Twitter, and street blocks, and natural cities from OpenStreetMap (OSM, Bennett 2010). Section 4 discusses our results using the London and Paris natural cities, although the results apply to all other natural cities with a single exception. Section 5 further discusses the implications of the study and its major findings in the context of big data. Finally, Section 6 draws conclusions and points to future work.

**2. Natural cities, street blocks and related distances**
The natural cities used in this paper are naturally and automatically delineated from a large amount of street blocks, usually all street blocks of a country, although natural cities could be defined with other big data such as social media location data and nighttime images (Jiang and Miao 2015, Jiang et al. 2015). In general, natural cities are defined as human settlements or human activities in general on Earth's surface that are objectively or naturally delineated from massive geographic information of various kinds, and based on the head/tail breaks – a relatively new classification for data with a heavy tailed distribution (Jiang 2013). The concept of natural cities was initially developed by Jiang and Liu (2012), in which the authors adopted street networks of the three European countries France, Germany, and the United Kingdom, and aggregate small blocks (smaller than an average) to constitute individual natural cities. These natural cities are used to examine the spatial distribution of tweets and densities in this study. For this purpose, in this section, we first illustrate the natural cities derived from the street blocks, and then present the topological center, and related distance concepts such as topological, and geometric, and radial.

A geographic space, such as a country, is considered to be composed of a large number of blocks: small blocks called city blocks, and large blocks called field blocks. Jiang and Liu (2012) developed an algorithm to automatically derive all the blocks and then select the small ones to form individual natural cities. This process is illustrated in Figure 1, in which a set of small blocks (smaller than an average) constitute the natural city indicated by the red border. The process of deriving natural cities is based on the head/tail division rule – the non-recursive version of the head/tail breaks (Jiang 2013). Given a variable X, if its values (xi) follow a heavy-tailed distribution, then we can put all the values xi into two unbalanced parts around an average: those greater than the average are called the head, which is a minority, and those less than the average are called the tail, which is a majority (Jiang and Liu 2012). As Figure 1 shows, the natural city or the city (or small) blocks as a whole occupy 5/16 = 31%, while the field (or large) blocks as whole occupy the other 69%. Natural cities present a powerful concept for studying city growth and evolution and for better understanding city structure and dynamics (Jiang and Miao 2015, Jiang 2015a).

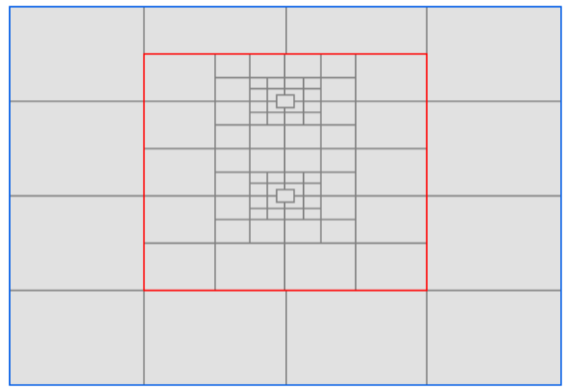

Figure 1: (Color online) Illustration of a natural city (within the red border) derived from the set of street blocks of a fictive country with the blue border
(Note: There are far more small (or city) blocks (70) than large ones or field blocks (12), and the city blocks as a whole occupy a small percentage of the area 5/16 =31%)

There are several concepts related to the street blocks perspective on a geographic space. For the sake of simplicity, we use the natural city in Figure 1 to illustrate these concepts (Figure 2). The border number indicates how far, in terms of topological distance, a street block is from the outmost border of a street network. All blocks on the border have border number 1, and all blocks adjacent to blocks with border number 1 have border number 2, and so on. Computing the border numbers (as well as the center numbers to be introduced) is very much like to find out one's friends, the friends of all the friends, and so on until all people are searched. The block or cell with the highest border number, or with the longest topological distance, is called the center, or more precisely the topological center. The topological center differs fundamentally from the geometric center, or centroid. For example, London and Paris are the topological centers of the UK and France, respectively, even though both cities are far from the geometric centers of the countries (Jiang and Liu 2012). Similar to the border number, the center number is defined as the topological distance from the center. Both the border and center numbers are derived using the 3×3 Moore neighborhood, which implies neighboring relationships shared by a side or a point. It should be noted that the border and center numbers are not inversely symmetric; the highest border number is usually smaller than the highest center number. For example, the highest border number of the city is 5, while the highest center number is 8 (Figure 2); the blocks with border number 1 are not those with the highest center number.

Given the center number or the topological distance from the center, we can define the corresponding geometric distance. Although it is computed from the corresponding topological distance, the fact that a block has the longest topological distance does not mean it has the longest geometric distance. For example, the red blocks in Panel b are no longer red in Panel c (Figure 2). The geometric distance appears to be quite consistent with our intuition regarding how far a street block is from the city center. Panel d of Figure 2 presents the commonly used radial distance, which is the direct Euclidean distance between the city center and a block. The radial distance looks very different from the topological and geometric distances. It is important to note that changing the size of the natural city would not affect topological distance (both the border and center numbers), but would significantly affect geometric and radial distances.

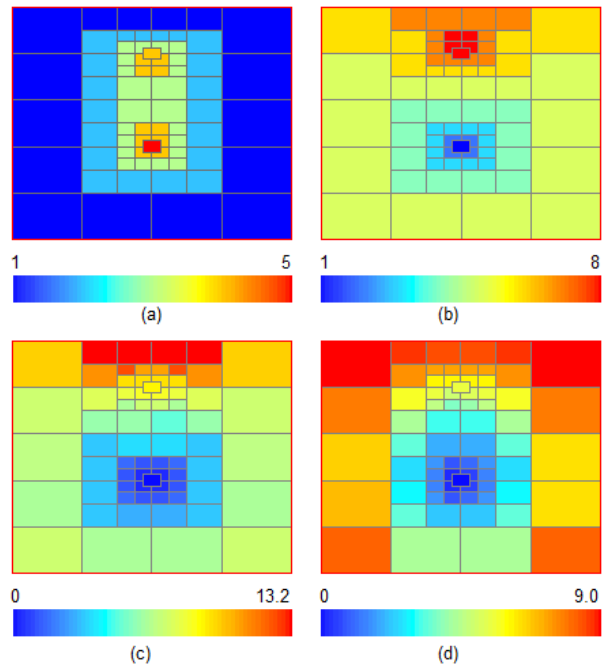

Figure 2: (Color online) Illustration of related distances based on city blocks of the natural city shown in Figure 1
(Note: The border number indicates how far topologically a block is from the outmost border (Panel a). The center number is the topological distance of a block from the center (Panel b). The center number can be transformed into the geometric distance, which is scale dependent (Panel c). The radial distance indicates the straight line distance between a block and the center (Panel d))

**3. Data and data processing**
We adopted street networks of France, Germany, and the United Kingdom for our study. Taken from OSM (Bennett 2010), the data have been processed previously as individual street blocks (Jiang and Liu 2012). It should be noted that the data is very big (up to 2 million blocks in the case of Germany). In this study, we relied on the street blocks data to extract individual natural cities, in total 30,000 natural cities, and then chose the two largest cities in each country (Paris, Toulouse, Berlin, Munich, London, and Birmingham) for reporting the results on spatial distribution of tweet numbers and densities. As a reminder, these are natural cities rather than conventional cities defined or imposed by authorities from the top down. To make the paper self-contained, we briefly introduce how the natural cities were obtained using a simple Python script and ArcGIS functionality.

All the blocks in any of the three countries follow a heavy-tailed distribution in terms of block areas. Given the heavy-tailed distribution, we used the average block size for each country to divide all the blocks into city blocks (smaller than the mean) and field blocks (greater than the mean). To derive the natural cities, a recursive function was developed to traverse each city block and cluster its adjacent city blocks whose adjacent blocks are also city blocks. In this way, we obtained the general extent of every natural city in a country. The pseudo-code of this clustering process looks as follows:

```
Recursive Function CityBlockClustering (start Block)
    If (this block is city block and is not processed)
       Add this block into Blocklist;
       Mark this block as processed;
       Get its adjacent blocks;
          Foreach adjacent blocks:
          If (its adjacent blocks are all city blocks):
             CityBlockClustering (this block);
    Return Blocklist;
End Function
```

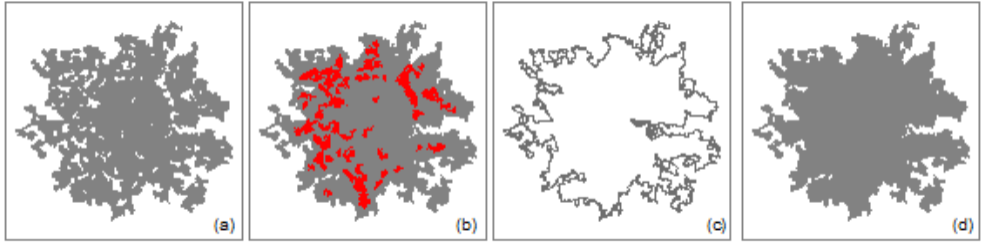

Figure 3: (Color online) Illustration of creating the London natural city
(Note: The creation process includes clustering the city blocks (a), filling the holes indicated by the red patches (b), obtaining the natural city boundary (c), and clipping out the natural city (d).)

After the above recursive function processing, all the city blocks are clustered into a connected whole, but with many holes inside. The next steps include filling the holes, and obtaining the boundary, and clipping out the natural city (Figure 3), which can be done with ordinary GIS tools. For the convenience of readers who are familiar with ArcGIS, we present a simple recipe for the process. First, the clustered city blocks are input to create the extent polygon by using Data Management Tools > Feature > Minimum Bounding Geometry (The option "Group" should choose "All"). Second, Analysis Tools > Overlay > Erase is followed to subtract the extent polygon and clustered city blocks to get the holes, which will further be merged with the city blocks (Analysis Tools > Overlay > Union). Subsequently, each natural city boundary can be obtained by dissolving the merged polygons through Data Management Tools > Generalization > Dissolve. Finally, all natural cities in a country can be extracted from the country blocks with the input of resulting natural city boundaries by following Analysis Tools > Extract > Clip. Following the above data processing, the six natural cities with very convoluted boundaries were obtained (Figure 4). Table 1 presents basic statistics for the six natural cities.

Table 1: Basic statistics for the six natural cities after the data processing

|  | Paris | Toulouse | Berlin | Munich | London | Birmingham |
|---|---|---|---|---|---|---|
| # of blocks | 67,497 | 13,844 | 30,101 | 27,555 | 45,627 | 14,855 |
| # of tweets | 1,032,545 | 16,124 | 41,253 | 20,730 | 405,367 | 50,126 |
| Highest border # | 41 | 13 | 24 | 27 | 29 | 16 |
| Highest center # | 104 | 44 | 82 | 90 | 96 | 58 |
| Area (sq. km) | 16,209 | 1,876 | 5,002 | 3,168 | 8,977 | 2,378 |

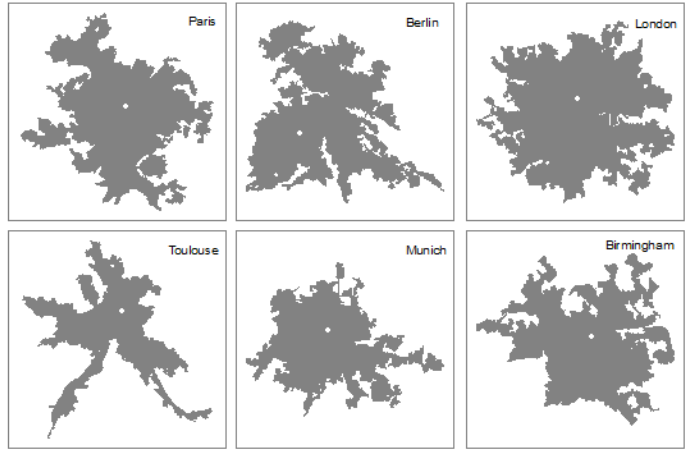

Figure 4: The six natural cities from the three European countries: France, Germany, and the UK
(Note: The white spot in the middle indicates the topological center)

We collected the tweet location data using the Twitter Stream API (Kumar et al. 2013) by specifying the region-of-interest represented by a bounding box. The data contain all tweets with GPS coordinates that fall in this bounding box. The Twitter streaming API has the 1% sample policy that the number of tweets released from the API should not exceed 1% of the total. Therefore, we made sure that there was no the kind of warning messages during the data collection about the data limit. The data collection time period for the UK is June 1–8, 2014, and for Germany and France it was September 17–24, 2014. The raw Twitter data collecting from the streaming API were converted into the location-related information, including user ID, latitude, longitude, and timestamp. The tweet numbers for the six natural cities are shown in Table 1. These tweet location data were assigned to individual street blocks in order to assess their spatial distributions (both tweet numbers and densities) from the city center to the periphery. The spatial distributions are further investigated with two what-if experiments: (1) what if census tracts are used as basic geographic units? and (2) what if an arbitrary imposed city boundary is adopted?

**4. Two major findings**
This study leads to two major findings. The first is that spatial distributions of tweets and densities are highly affected by geographic units, but less so by different types of distances. From the city center to the periphery, the tweets first increase and then decrease, while the densities decrease in general; see Figures 5 and 6 for London and Paris, respectively. On the one hand, these two spatial variations appear to be quite consistent with respect to the three different distances: topological, geometric, and radial. On the other hand, the spatial variations fluctuate dramatically, somewhat like stock prices during a time period. To compare, the fluctuation does not appear or is less significant, when all the tweet locations are assigned into the 628 London census tracts (obtained from the Ordnance Survey OpenData program) (Figure 7). Instead, the spatial variations look quite smooth and are very similar to urban population densities in the literature (e.g., Clark 1951, Wang and Zhou 1999, Chen 2009). In this regard, we believe that these previous studies relied on some coarse geographic units such as census tracts, and therefore reached the conclusion that urban population densities follow exponential or power law functions. The first finding is commonly known as the modifiable areal unit problem (MAUP, Openshaw 1984).

The MAUP is a well-recognized problem in spatial analysis because different geographic units are used for aggregating point-based data. The MAUP was first discovered by Gehlke and Biehl (1934) and was popularized by Openshaw (1984). In the present study, we have seen clearly how the fine scale of city blocks (Figure 5) leads to significantly different spatial patterns against the coarse scale of census tracts (Figure 7). The MAUP is considered as a source of statistical biases or errors (Fotheringham 1989, Tobler 1989). In fact, it is an inevitable and unsolvable problem, and the best way to deal with it is to use the finest units, if possible, and to be aware of it when the coarse units are used. Based on the plots in Figures 5 and 7, we argue that the MAUP arises from the fractal nature of geographic space, and is the same nature of problem as the conundrum of length. The length of geographic features such as coastlines depends on map scales; the larger the map scale, the longer the coastlines (Richardson 1961). By plotting the spatial distribution of tweets, we actually transformed the MAUP into the problem of conundrum of length. The spatial distribution curves in Figure 5 are much more convoluted and longer than those in Figure 7; see a further discussion in the next section.

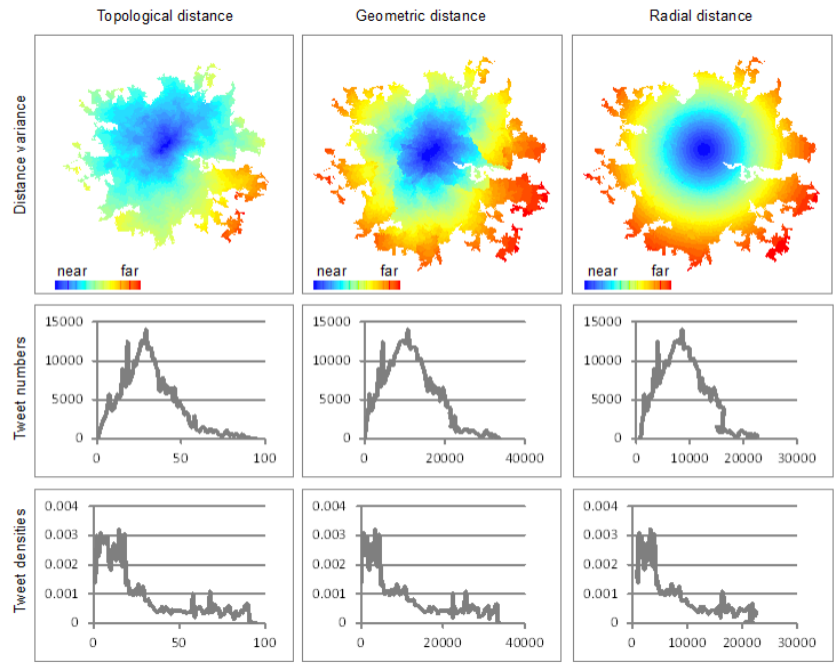

Figure 5: (Color online) Spatial distribution of tweets and densities from the center of London to the periphery based on the street blocks

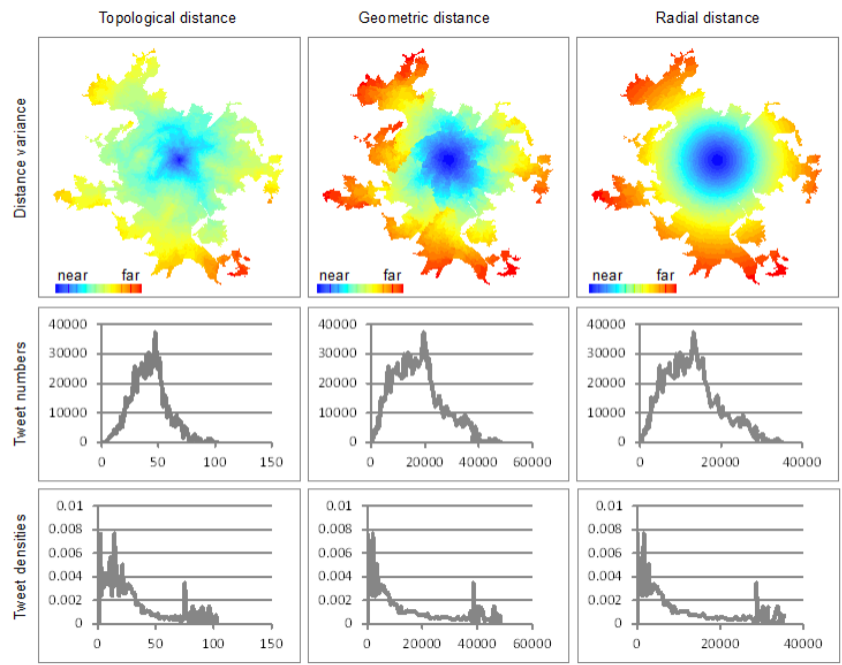

Figure 6: (Color online) Spatial distribution of tweets and densities from the center of Paris to the periphery based on the street blocks

The second major finding is that the observed spatial distributions of tweets and densities are less significant, or simply disappear, when an arbitrarily imposed city boundary is adopted. To illustrate, we adopted a central part of Paris surrounded by a ring road. As shown in Figure 8, the spatial distributions are strikingly different from those in Figures 5 and 6. This is understandable, since only part of Paris is used for showing the spatial distributions. This finding indicates the power of natural cities, or equivalently, the naturally delineated city boundaries, in geographic research. If all the street blocks of a country are considered as a whole, then those street blocks of a natural city constitute a sub-whole. We believe the natural cities or their boundaries quite accurately reflect what are commonly perceived as cities. This boundary effect is in fact the uncertain geographic context problem (UGCoP, Kwan 2012), which will be further discussed below.

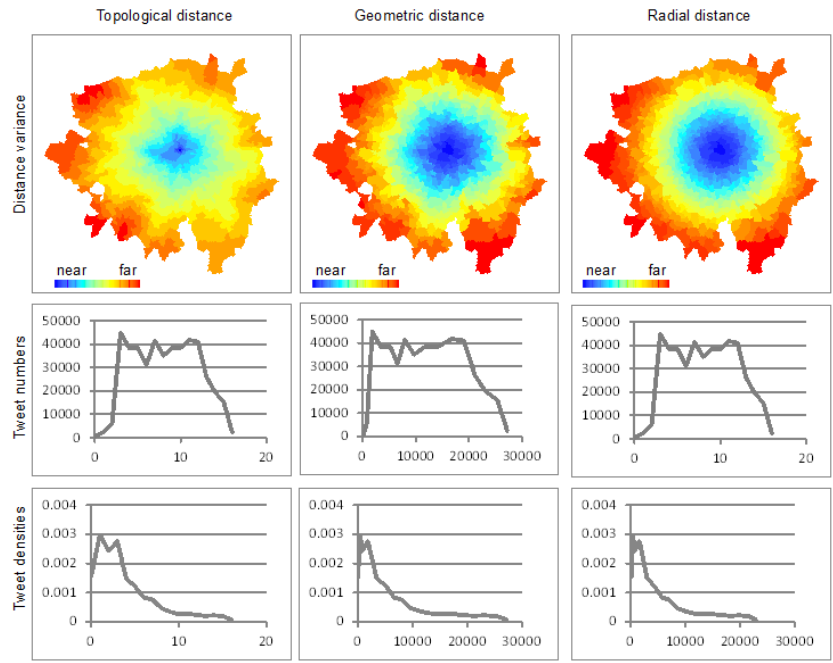

Figure 7: (Color online) Spatial distribution of tweets and densities from the London center to periphery based on the census tracts

A natural city boundary should look like a wiggly coastline (Figure 4) rather than a regular or smooth shape (Figure 8). This is the same for many country boundaries in the world. The present study developed a holistic view of delineating appropriate units for spatial analysis. From a large amount of street blocks of a country, we delineated the natural cities following the head/tail breaks and considering spatial auto-correlation effect. The natural city boundaries enable us to automatically and naturally detect their centers in order to further study spatial distributions of tweets from city centers to peripherals. In this connection, the natural cities act as sub-wholes derived from all of the street blocks of a country as a whole. We have seen through the study that natural cities are appropriate geographic units or contexts for studying spatial distributions of tweets. We believe that the way of delineating the natural cities sets a good example for dealing with the UGCoP (Kwan 2012). Further research is needed to apply the notion of natural cities at a city scale, in which districts, neighborhoods, and hotspots would be delineated to constitute sub-wholes at different levels.

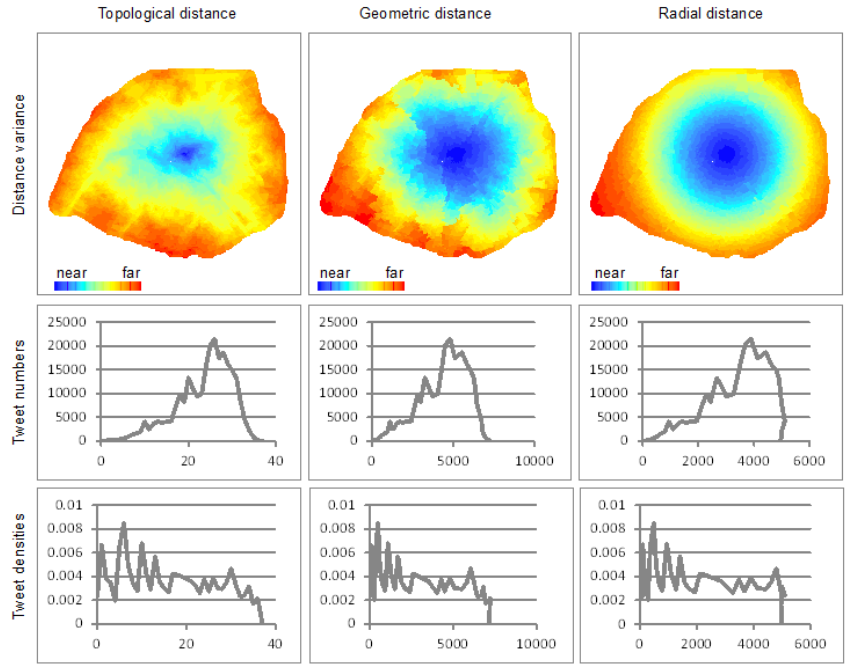

Figure 8: (Color online) Spatial distribution of tweets and densities from the Paris center to the periphery based on an arbitrarily imposed city boundary

London and Paris were chosen for illustration, but the two findings apply fairly well to all of the other 4 natural cities, as well as 30,000 natural cities from the three countries. However, Berlin is probably the only exception among all the natural cities in the three countries. The natural city of Berlin does not show consistent results with other natural cities. This is mainly because the auto-detected topological center is dramatically deviated from the real city center of Berlin called Mitte. The Berlin natural city boundary looks odd; in particular, the east part of the city is very fragmented and concave. The oddity is attributed to the fact that the east part is mixed with a river, rail lines, and green parks that lead to some big blocks. Another possible reason could be that East Berlin is less developed due to the Cold War, which led to some blocks being larger than the average. We checked all other natural cities, but did not notice such an oddity.

**5. Implications of the study and its findings**
Large amounts of social media data enable us to study human activities on very fine spatial and temporal scales (e.g., Jiang and Miao 2015). Instead of aggregating the data into existing geographic units such as census tracts, we assigned individual locations into auto-detected city blocks to study the spatial distribution of tweets in cities. The spatial distributions from the city centers to the peripheries look very much like fluctuated stock prices over a period of time. The fluctuation or roughness is the essential nature of geographic phenomena (e.g., Goodchild and Mark 1987, Batty and Longley 1994), and indeed of many other natural and societal phenomena (Mandelbrot 1982, Mandelbrot and Hudson 2004, Bak 1996). In this section, we further discuss the implications of this study and its major findings in the context of big data.

The natural cities derived from large amounts of street blocks can be considered as an emergence of the complex street networks. The diverse and heterogeneous blocks collectively determine who is with whom to constitute the natural cities, and who else can be rural areas. The block sizes are clearly right-skewed in the histogram or heavy-tailed in the rank-size plot (Zipf 1949). Given the heavy-tailed distribution, and according to the head/tail division rule (Jiang and Liu 2012), the small and large blocks constitute the natural cities and rural areas, respectively. The heavy-tailed distribution of the blocks is what drove us to develop the notion of natural cities. The street blocks are strongly auto-correlated, which makes the delineation of the natural cities possible. To paraphrase Tobler's (1970) first law of geography, every block is related to every other block, but near blocks are more related than distant blocks. In this sense, the natural cities are an outcome of the first law of geography. The method of deriving the natural cities can be extended to a city scale, in which we would be able to derive districts, neighborhoods or hotspots in cities. This, of course, warrants further studies.

Every block is related to every other block, and this relatedness can be seen from the border and center numbers, or equivalently, the topological distances far from the border and center. Taking the center number, for example, every block is related to the city center blocks, but only those directly surrounding the center blocks are the most related ones. In this connection, the topological distance based on street blocks indicates the degree of the relatedness: the shorter the distance, the more related the street blocks. The notion of relatedness is what motivated us to define the topological distance rather than the conventionally used radial distance. Readers may argue that, with respect to Figures 5 and 6, the topological distance does not show much difference from the radial distance. While this is true, the differences among the three distances are clearly shown in Figure 2. More importantly, it is the topological distance or the border number that enables us to identify the city center – the blocks with the highest border number.

Near blocks are more related than distant blocks, and this nearness can be better measured by the topological distance than the radial distance. All the three distances are essentially geographical, despite their differences. The notion of nearness can also be understood from a statistical perspective, or the rank-size plot (Zipf 1949). If all the blocks are ranked from the largest to the smallest, this ranking series constitutes a one-dimensional space. Interestingly, the notion that near blocks are more related than distant blocks applies to this one-dimensional space as well; those big blocks in the head are more related (being rural), whereas the small blocks in the tail are more related, forming the individual natural cities. If all the natural cities are ranked in the same way as the blocks, we would find that near things in the ranking series are indeed more related than distant things. For example, the city of New York is more related to Los Angeles than a geographically nearby town or village, because the two cities are the two largest in the United States.

The fluctuation of tweet numbers and densities looks very much like the daily stock prices seen in financial newspapers. There is little doubt that stock prices, as well as many other natural and societal phenomena, are fractal (Mandelbrot and Hudson 2004, Bak 1996). The development of fractal geometry (Mandelbrot 1967) was initially inspired by the connundrum of length (Richardson 1961). While geographers (Nystuen 1966, Perkal 1966) tried to solve the connundrum of length in order to measure things in maps, the mathematician developed fractal geometry (Mandelbrot 1982). It is well recognized that the connundrum of length arises out of the fractal nature of geograpic features. It is the same for the MAUP. The curves in the plots look rough, and wiggly because of the fine street blocks (Figures 5 and 6), yet they look smooth or less wiggly because of the coase census tracts (Figure 7). Given the rough nature, fractal geometry must become the most appropriate paradigm for geographic research, particularly in the big data era. This is the same in spirit as calling for Paretian thinking for geospatial analysis (Jiang 2015b), since in the big data era things are more likely to be scale free, no well-defined mean.

We conjecture that the spatial distributions of tweets illustrated in this paper quite accurately reflect those of urban populations. In other words, tweet densities are a good surrogate of population densities. One could argue that social media are biased towards certain groups, such as young people or those with Internet access. While this is true, the point is that the massive amount is big enough, or heterogenerous enough to be characterized by all, both literally and metaphorically. For example, it is estimated that only one percent of all tweets have GPS locations that can be georeferenced, and others are just place names. However, this one percent is already large enough. For example, several million tweets are made with GPS locations around the world in a 24-hour period. These tweet locations can fairly accurately characterize human activities across the globe, despite the large holes in countries such as China and North Korea. This conjecture deserves further research in the future.

# 6. Conclusion

This study investigated the spatial distributions of tweets in cities at the scale of city blocks for natural cities, together with two what-if experiments: (1) what if the tweets are assigned into census tracts rather than city blocks? and (2) what if we used an arbitrarily imposed city boundary rather than a natural city? We found that, from the city center to the periphery, tweets first increase and then decrease, and that the tweet densities decrease in general with respect to three different distances: topological, geometric, and radial. However, both the increase and decrease fluctuated dramatically. This fluctuation disappears or is less significant if census tracts or an arbitrarily imposed city boundary are used. These two findings have some profound implications for geographic research or spatial analysis in the big data era. The fluctuated spatial distributions are the true picture of geography – the fractal or scaling property of natural and man-made phenomena. In this connection, social media data, or big data in general, provide a new instrument for geographic research, and enable us to better understand the underlying fractal structure and nonlinear dynamics.

This study has shown the power of natural cities in geographic research, particularly in the big data era when we face large amounts of location-based data emerging from the Internet. We have also seen the advantages of topological distance based on street blocks. It enables us to derive the true center of a geographic space and helps us to assess the relatedness and nearness with respect to the first law of geography. The first law of geography was reformulated for a large amount of street blocks, as well as natural cities, in the one-dimensional ranking space from the largest to the smallest. We have argued that the MAUP is essentially an inevitable problem, arising from the fractal nature of geographic features, but big data and the finest geographic units help better deal with the problem. We further conjecture that tweet densities can be a good surrogate of population densities in cities. All these issues merit further investigation in the future.

**Acknowledgement**

We would like to thank Dr. Waldo Tobler and Dr. Daniel Sui for their constructive comments on an earlier and later version of this paper. This research is partially supported by special fund of Key Laboratory of Eco Planning & Green Building, Ministry of Education (Tsinghua University), China.